\documentclass[11pt]{article}
\usepackage[letterpaper, total={6in, 9in}]{geometry}

\usepackage{setspace}

\usepackage{cite}
\usepackage[pdftex]{graphicx}
\graphicspath{{pdf/}{jpeg/}{figures/}}
\DeclareGraphicsExtensions{.pdf,.jpeg,.png}
\usepackage{url}

\usepackage{amsmath}

\begin{document}
\title{Compressed Sensing: From Research to Clinical Practice with Data-Driven Learning}
\author{
    Joseph~Y.~Cheng\thanks{J.~Y.~Cheng is with the Department
of Radiology, Stanford University, Stanford, CA, 94022 USA e-mail: jycheng@stanford.edu.},
    Feiyu~Chen,
    Christopher~Sandino,
    Morteza~Mardani,\\
    John~M.~Pauly,~\textit{Senior Member,~IEEE,}
    Shreyas~S.~Vasanawala\thanks{F.~Chen, C.~Sandino, M.~Mardani, J.~M.~Pauly, and S.~S.~Vasanawala are with Stanford University.}}
\date{}
\maketitle

\begin{abstract}
Compressed sensing in MRI enables high subsampling factors while maintaining diagnostic image quality. This technique enables shortened scan durations and/or improved image resolution. Further, compressed sensing can increase the diagnostic information and value from each scan performed. Overall, compressed sensing has significant clinical impact in improving the diagnostic quality and patient experience for imaging exams. However, a number of challenges exist when moving compressed sensing from research to the clinic. These challenges include hand-crafted image priors, sensitive tuning parameters, and long reconstruction times. Data-driven learning provides a solution to address these challenges. As a result, compressed sensing can have greater clinical impact. In this tutorial, we will review the compressed sensing formulation and outline steps needed to transform this formulation to a deep learning framework. Supplementary open source code in python will be used to demonstrate this approach with open databases. Further, we will discuss considerations in applying data-driven compressed sensing in the clinical setting. 
\end{abstract}

\section{Introduction}
Compressed sensing is a powerful tool in magnetic resonance imaging (MRI). As the scan duration is directly related to the number of data samples measured, collecting fewer measurements enables faster imaging. Multi-channel imaging, also known as ``parallel imaging,'' leverages the localized sensitivity profiles of each receiver element in a coil array to enable subsampling factors on the order of 2--6 \cite{Pruessmann1999,Ying2007,Uecker2013}. By exploiting properties of the reconstructed images (such as the sparsity of the Wavelet coefficients of an image), compressed sensing can further increase the subsampling factors by 2--3 fold \cite{Lustig2007a}. The additional factor is extremely powerful in enabling a broad range of clinical applications. For example, high-resolution volumetric imaging that would take minutes to acquire can be achieved in a single breathhold. This strategy minimizes the sensitivity of patient motion and increases the diagnostic quality of the resulting images.

\begin{figure}[ht]
    \centering
    \includegraphics[width=0.7\textwidth]{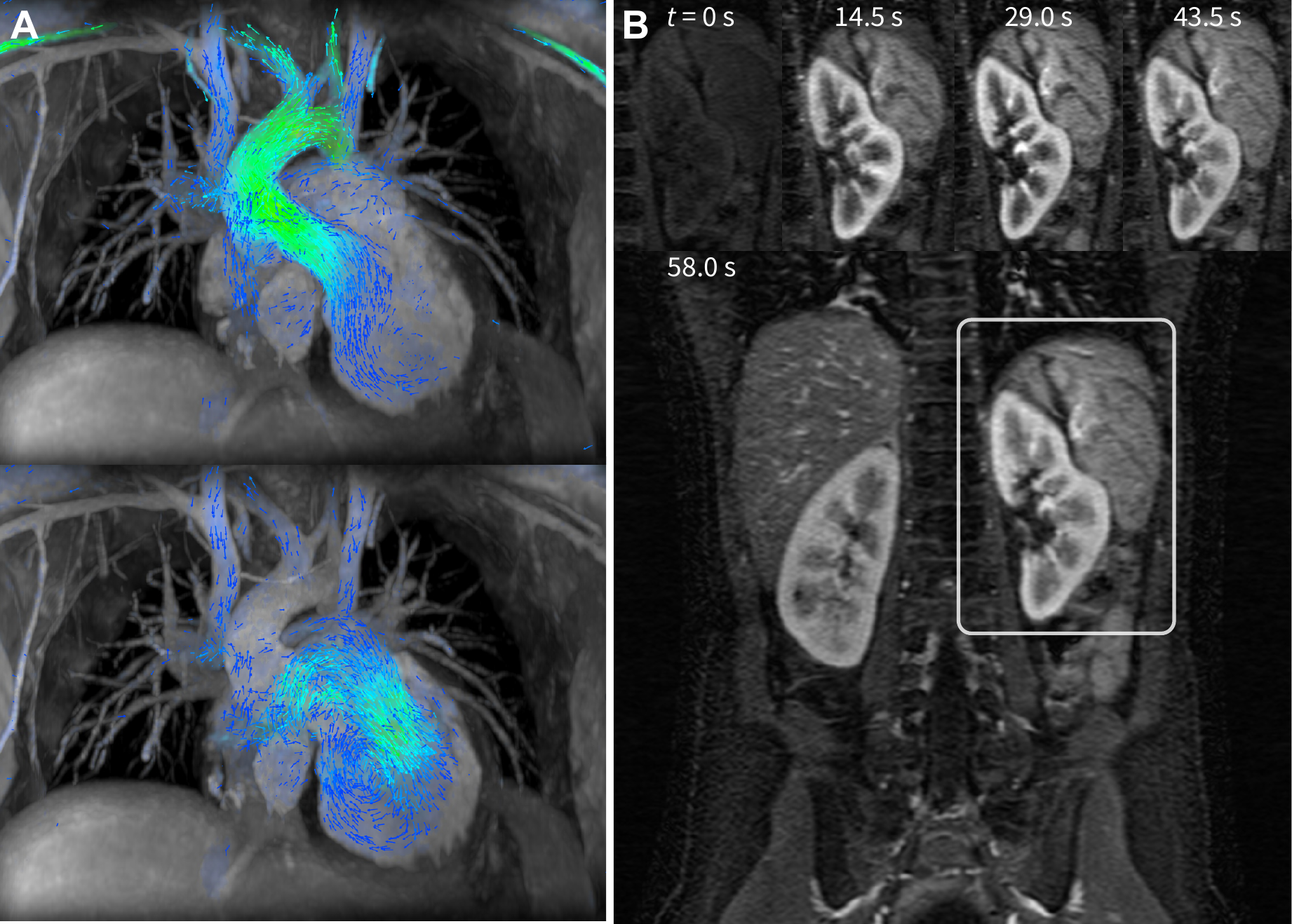}
    \caption{Example clinical applications of compressed sensing. In \textbf{A}, two cardiac phases of the cardiac-resolved volumetric velocity MRI (4D flow) are shown for the purpose of congenital heart defect evaluation \cite{Vasanawala2015}. To enable a clinically feasible scan duration of 5--15 minutes while maintaining high spatial ($0.9\times0.9\times1.6$ mm$^3$) and high temporal resolutions (22.0 ms), a subsampling factor of 15 was used \cite{Cheng2015a}. In \textbf{B}, high spatial ($1\times1\times2$ mm$^3$) with a 14.5-s temporal resolution was achieved for a dynamic contrast enhanced MRI using an acceleration factor of 6.5 \cite{Zhang2013a}. High spatial-temporal resolutions are required for capturing the rapid hemodynamics of pediatric patients.}
    \label{fig:example}
\end{figure}

Over the past decade since compressed sensing was introduced to MRI \cite{Lustig2007a}, many developments have been made to extend this idea and bring it into clinical practice. One area with significant research and clinical activity is in multi-dimensional imaging. With more dimensions to exploit sparsity, the subsampling factor can be increased substantially (over 10 fold). As a result, multi-dimensional scans can be completed in clinically feasible scan times. For example, a volumetric cardiac-resolved flow imaging sequence (4D flow) can be performed in a 5--10 min scan instead of an hour long scan needed to satisfy the Nyquist rate \cite{Cheng2015a}. This single 4D flow scan enables a comprehensive cardiac evaluation with flow quantification, functional assessment, and anatomical information \cite{Vasanawala2015}. Instead of an hour long cardiac exam for congenital heart defect patients with complex cardiac anomalies, the exam can be completed in a simple-to-execute single 4D flow scan (Fig.~\ref{fig:example}A). Other examples of multi-dimensional compressed sensing include cardiac imaging 
\cite{jung2009k,Feng2012}, 
dynamic-contrast-enhanced imaging \cite{Lingala2011,Zhang2013a} (Fig.~\ref{fig:example}B), and ``extra-dimensional'' imaging with 4+ dimensions
\cite{Feng2015,Cheng2017ScientificReports}.

Additionally, rapid imaging has significant impact on the clinical workflow. Exam times can be significantly shortened to reduce patient burden and discomfort. For pediatric imaging, the shortened exam time enables the reduction of the depth and length of anesthesia \cite{Vasanawala2010a}. For extremely short scan times (less than 15 minutes), anesthesia can entirely eliminated.

\subsection{Remaining Challenges}
\begin{figure}[ht]
    \centering
    \includegraphics[width=0.8\textwidth]{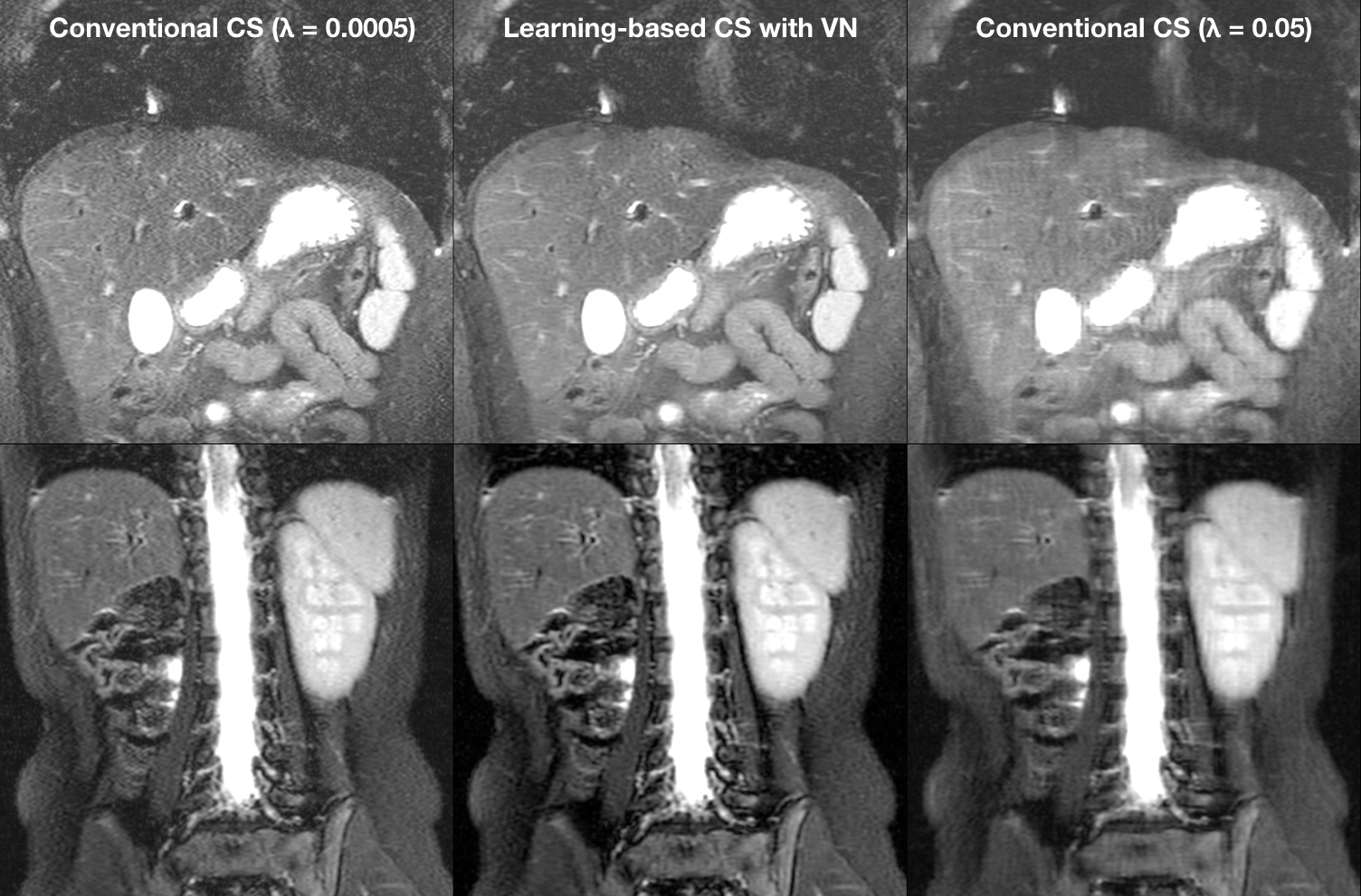}
    \caption{Example results of data-driven compressed sensing (CS) with variational networks (VN) \cite{Hammernik2017,Chen2018b}. Conventional compressed sensing requires tuning of the regularization parameter $\lambda$. The optimal value of this parameter may vary with different scans. Compared to conventional CS using L1-ESPIRiT \cite{Uecker2013} (left and right columns), learning a VN (middle column) achieves proper regularization without the need of manual tuning the regularization parameter.}
    \label{fig:demo:cs_reg}
\end{figure}

Much success has been observed by applying compressed sensing to specific clinical applications such as for cardiac cine imaging or for MR angiography. However, the potential that compressed sensing brings is not fully realized in terms of maximal acceleration and breadth of applications. A number of challenges limit the use of compressed sensing for clinical practice. First, compressed sensing is sensitive to tuning parameters. One example of these parameters is the regularization parameter. These parameters determine the weight of the regularization term with respect to the data consistency term in the cost function. Increasing the regularization parameter will improve the perceived signal-to-noise ratio (SNR) of reconstructed images. However, while the perceived SNR improves, fine structures in the reconstructed images may be over-smoothed, or new image textures may be introduced, resulting in an artificial appearance of images. The optimal value of the regularization parameter usually varies among scans and patients, making parameter tuning clinically infeasible. The impact of the regularization parameter on the reconstruction performance is illustrated in Fig.~\ref{fig:demo:cs_reg}.

Second, compressed sensing may result in textural artifacts. Compressed sensing theory assumes that the unknown signal is sparse in some transform domain. This domain is often manually chosen based on the characteristic of the image. For MRI applications, common choices of this domain transformation are wavelet \cite{Lustig2007a}, total variation \cite{Feng2012}, and locally low-rank transforms \cite{Lingala2011}. However, these transforms may promote some textural artifacts in the reconstructed images. For example, locally low-rank reconstruction may cause block artifacts in reconstructed images.

Lastly, compressed sensing usually has unknown reconstruction times because of the use of iterative convex optimization algorithms. To achieve clinically acceptable image quality, the number of iterations for these algorithms may vary between 50 to over 1000, resulting in varying reconstruction times. For volumetric scans, these numbers of iterations correspond to several minutes to over an hour long reconstruction times. This uncertainty of long reconstruction times may further lead to delays and queues in clinical scanning, as well as require expensive dedicated computational hardware.

\subsection{Current Solutions}
Much research has attempted to solve these challenges. The performance of compressed sensing is highly dependent on the regularization function and regularization parameters used. To reduce textural artifacts, domain transformations used in the regularization terms (i.e. Wavelet transform, total variation) are carefully chosen and tuned for each application. Impact of these different regularization functions must be considered as different functions introduce different types of bias to the reconstructed images.

Ideally, the regularization parameters should be set based on the system noise. However, many other factors such as patient motion contribute to measurement error. Thus, regularization parameters are often manually tuned for each clinical application to achieve acceptable performance. Automated parameter selection methods, such as Stein's unbiased risk estimate (SURE) \cite{ramani2012regularization}, have been developed to select the regularization parameters for compressed sensing MRI. However, this approach comes at the cost of additional reconstruction times; as a result, this approach is uncommon in practice. In regards to reconstruction time, a trade-off between reconstruction time and performance can also be achieved by early truncation of the iterative reconstruction \cite{Beck2009}.

Most of these solutions require manual adjustment of the original compressed sensing problem to specific applications. This tuning and redesigning process requires extra time and effort, and constant attention as hardware and clinical protocols evolve. Previous solutions to automate some of this process involve additional memory and more complex computations.

\subsection{Overview}
To overcome the obstacles with compressed sensing, data-driven learning has recently become a compelling and practical approach
\cite{Yang2016a,Hammernik2017,Adler2017,aggarwal2018modl,Cheng2018}.
The purpose of this tutorial is to build the basic framework for extending compressed sensing to a data-driven learning approach and to describe the considerations for clinical deployment. We also discuss a number of new challenges when using this framework. We have released supplementary python code on GitHub\footnote{\url{https://github.com/MRSRL/dl-cs}} to demonstrate an example of this data-driven compressed sensing framework.

\section{Data-Driven Compressed Sensing} 
\subsection{Background}
The MRI reconstruction problem can be formulated as a minimization problem \cite{Lustig2007a}. The optimization consists of solving the following equation:
\begin{equation}
    \hat{m} = \arg\min_m \frac{1}{2}\|Am - y \|_2^2 + \lambda R(m)\label{eq:recon}
\end{equation}
where $m$ is the reconstructed image set, $A$ describes the imaging model, and $y$ is the measured data in the k-space domain. The imaging model for MRI consists of signal modulation by coil sensitivity profile maps $S$, Fourier transform operation, and data subsampling. These sensitivity profile maps $S$ are specific for each dataset $y$. 

The goal of this optimization is to reconstruct an image set $m$ that best matches the measured data $y$ in the least squares sense. For highly subsampled datasets, this problem is ill posed -- many solutions satisfy Eq.~\ref{eq:recon}; thus, the regularization function $R(m)$ and corresponding regularization parameter $\lambda$ incorporate image priors to help constrain the problem. Many optimization algorithms have been developed to solve the minimization problem in Eq.~\ref{eq:recon}.
For simplicity, we base our discussion on the proximal gradient method. We refer the reader to similar approaches based on other optimization algorithms \cite{Yang2016a,Adler2017,Hammernik2017,Cheng2018}.

To solve Eq.~\ref{eq:recon}, we split the problem into two alternating steps that are repeated. For the $k$-th iteration, a gradient update is performed as 
\begin{equation}
    m^{(k+)} = m^{(k)} - 2tA^H (A m^{(k)} - y),\label{eq:grad}
\end{equation}
where $A^H$ is the transpose of the imaging model, and $t$ is a scalar specifying the size of the gradient step. The current guess of image $m$ is denoted here as $m^{(k+)}$. Afterwards, the proximal problem with regularization function $R$ is solved:
\begin{equation}
    m^{(k+1)} = \textrm{\textbf{prox}}_{\lambda R}\left( m^{(k+)} \right) = \arg\min_u R(u) + \frac{1}{2\lambda}\|u - m^{(k+)}\|_2^2,\label{eq:prox}
\end{equation}
where $u$ is a helper variable that transforms the regularization into a convex problem that can be more readily solved. The updated guess of image $m$ at the end of this $k$-th iteration is denoted as $m^{(k+1)}$, and $m^{(k+1)}$ is then used for the next iteration in Eq.~\ref{eq:grad}. This proximal problem is a simple soft-thresholding step for specific regularization functions such as $R(m)=\|\Psi m\|_1$ where $\Psi$ is a Wavelet transform \cite{Daubechies2004a}.

Previously, this regularization function has been hand-crafted and hand-tuned for every specific application. For example, spatial wavelets is a popular choice for general two and three dimensional images \cite{Lustig2007a}, spatial total variation (or finite differences) for angiography \cite{Lustig2007a}, and temporal total variation or sparsity in the temporal Fourier space for cardiac motion \cite{Feng2012}. Unfortunately, the design and testing of different regularization functions require significant engineering effort, and its practicality is hampered by the need to empirically tune the associated regularization parameters.

\subsection{Data-Driven Learning}
To be able to develop fast and robust reconstruction algorithms for different MRI sequences and scans, a compelling alternative is to take a data-driven approach to learn the optimal regularization functions and parameters. Though it may be possible to directly learn this regularization function, a simpler and more straight-forward approach is to learn the proximal step in Eq.~\ref{eq:prox} which will implicitly learn both the regularization function and regularization parameter. In this setup, the proximal step is replaced with a deep neural network to be learned: $E_{\theta_k}$ where $\theta_k$ are the learned parameters. Eq.~\ref{eq:prox} becomes
\begin{equation}
    m^{(k+1)} = E_{\theta_k}(m^{(k+)}).\label{eq:deeprox}
\end{equation}

\begin{figure}[ht]
    \centering
    \includegraphics[width=0.8\textwidth]{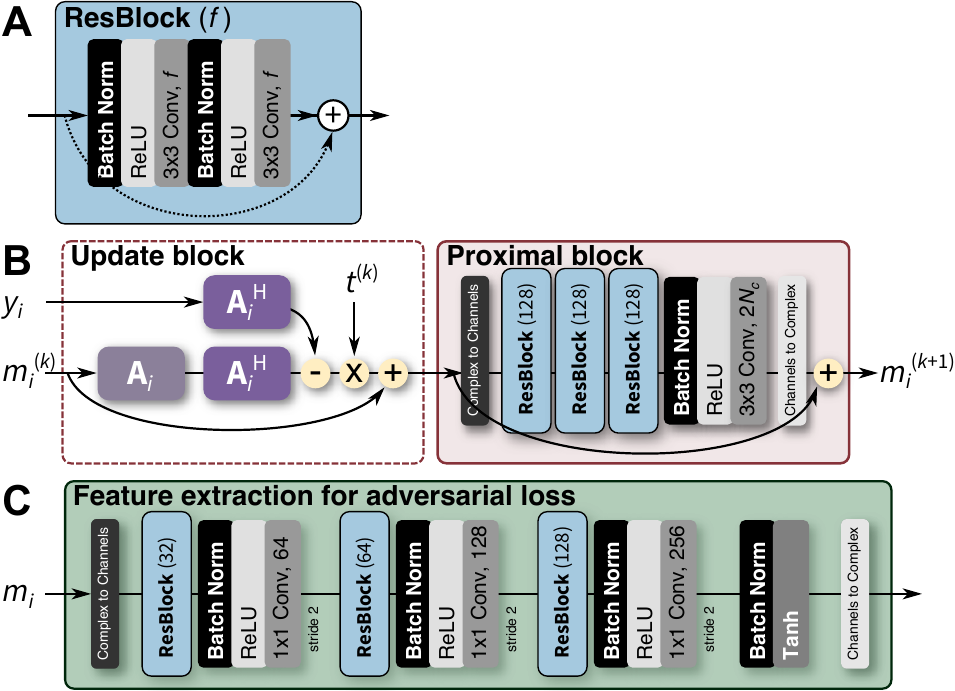}
    \caption{Neural network architecture for MRI reconstruction. One residual block (ResBlock) \cite{He2016a} is illustrated in \textbf{A} with $f$ feature maps. The ResBlock is used as a building block for the different networks. In \textbf{B}, one iteration of the reconstruction network is depicted where the $i$-th dataset is passed through the $k$-th iteration. Matrix $A_i$ represents the imaging model, $m_i$ represents the dataset in the image domain, and $y_i$ represents the dataset in the k-space domain. The final output can be passed through the network in \textbf{C} to extract feature maps that can be compared to the feature maps extracted from the ground truth data using the same network. The $\tanh$ activation function in \textbf{C} is used to ensure that the values in the outputted feature maps are within $\pm1$.}
    \label{fig:network}
\end{figure}

The steps of Eqs.~\ref{eq:grad} and \ref{eq:deeprox} can be unrolled with a fixed number of iterations and be denoted as model $G_\theta(y, A)$ with inputs of measurements $y$ and imaging model $A$. The training of such a model can then be performed using the following loss function:
\begin{equation}
    \min_\theta \sum_i \|G_\theta(y_i, A_i) - m_i\|_2^2,\label{eq:training}
\end{equation}
where $m_i$ is the $i$-th ground truth example that is retrospectively subsampled by a sampling mask $M$ in the k-space domain to generate $y_i = M A_i^H m_i$. For deployment, new scans are acquired according to the sampling mask $M$, and the measured data $y_i$ can be used to reconstruct images $\hat{m}_i$ as
\begin{equation}
    \hat{m}_i = G_\theta(y_i, A_i),
\end{equation}
where $A_i$ contains sensitivity profile maps $S_i$ that can be estimated using algorithms like JSENSE \cite{Ying2007} or ESPIRiT \cite{Uecker2013}. An example network is shown in Fig.~\ref{fig:network}AB.

\subsection{Neural Network Design for MRI Reconstruction}

Important considerations must be made when applying deep neural networks to MRI reconstruction. This includes handling of complex data \cite{Trabelsi2017}, circular convolutions in image domain for Cartesian data \cite{Cheng2018}, incorporating the acquisition model, flexibility in acquisition parameters and geometry, data normalization strategies, and plausible data augmentation approaches.

Since the data measured during an MRI scan is complex, complex data is used for the MRI reconstruction process. As a result, the networks used for MRI reconstruction need to handle complex data types. Two approaches can be used to solve this issue. The first approach is to convert the complex data into two channels of data. This conversion can be performed as concatenating the magnitude of the data with the phase of the data in the channels dimension, or concatenating the real component with the imaginary component of the data in the channels dimension. This conversion can be reversed without loss of data. The hypothesis is that the complex data properties may not need to be fully modeled with sufficiently deep networks. This first approach requires no modifications to current deep learning frameworks that do not have complex number support; however, the known structure of complex numbers are not fully exploited. The second approach is to build the neural network with operations that support complex data. Several efforts have been made to enable complex operations in the convolutional layers (with complex back-propagation) and the activation functions \cite{Trabelsi2017}. For demonstration purposes, we construct the networks in Fig.~\ref{fig:network}BC with the first approach because of its simplicity.

\begin{figure}
    \centering
    \includegraphics{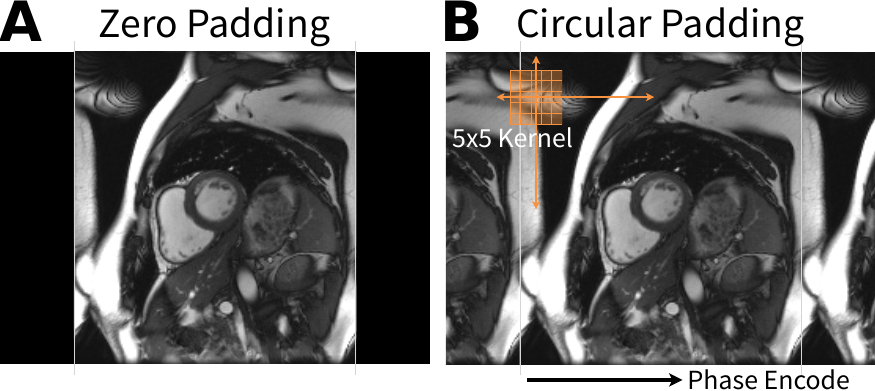}
    \caption{Circular convolutional layer for MRI reconstruction. Current deep learning frameworks support zero-padding for convolutional layers (shown in \textbf{A}) to maintain the original dimensions. However, data are measured in the frequency space, and the FFT algorithm is used to convert this data into the image domain. Thus, for Cartesian imaging, circular convolutions should be performed. This circular convolution can be performed by first circularly padding the input data as shown in \textbf{B} and cropping the result to the original input dimension.}
    \label{fig:circconv}
\end{figure}

Another consideration in building neural networks for MRI reconstruction is handling the convolutional operation at the image edges. Assuming zeros beyond the image edges for image-domain convolutions is sufficient for most cases, especially when the imaging volume is surrounded by air and when no high-intensity signal exists near the edges of the field-of-view (FOV). However, when these conditions are not satisfied, this assumption may result in residual aliasing artifacts. These artifacts are usually observed when the imaging object is larger than the FOV. Data are measured in the Fourier space and transformed using the fast Fourier transform algorithm (FFT) to the image domain; as a result, signals are circularly wrapped. Thus, circular convolutions should be performed when applying convolutions in the image domain. For this purpose, we first circularly pad the images before applying the convolutional layers (Fig.~\ref{fig:circconv}B). The padding width is set such that the ``valid'' portion of the output has the same image dimensions as the input.

Leveraging a more accurate MRI acquisition model in the neural network will improve the reconstruction performance and simplify the network architecture. More specifically, data are measured using multi-channel coil receivers. The sensitivity profile maps of the different coil elements can be leveraged as a strong prior to help constrain the reconstruction problem. Further, by exploiting the (soft) SENSE model \cite{Pruessmann1999,Uecker2013}, the multi-channel complex data ($y$) are reduced to a single-channel complex image ($m$) before each convolutional network block denoted as the proximal block in Fig.~\ref{fig:network}B. As a further benefit, the learned network can be trained and applied to datasets with different numbers of coil channels, because the input to the learned convolutional network block only requires a single-channel complex image.

Acquisition parameters may change for different patients, including spatial resolution, matrix size, and image contrast. Therefore, when applied to clinical scans, the neural networks need to be able to deal with these different scan parameters that are determined by each specific clinical application. One common parameter is the acquisition matrix size. Convolutional layers are flexible to different input sizes. For layers that are strict in sizes, the input k-space can be zero-padded to the same size to make the learned networks flexible to different dimensions of the acquisition matrix.

Appropriate data normalization helps improve the training and the final performance of the learned model. The input data should ideally be pre-whitened and normalized by an estimate of the noise statistics among the different coil array receivers. In the training loss function of Eq.~\ref{eq:training}, this data whitening helps balance the training examples based on the signal-to-noise ratio of each training example. These statistics can be measured during a fast calibration scan during which data are measured with no RF excitation. Alternatively, if this noise information is not available such as for already collected datasets, the noise can be estimated from the background signal for a fully sampled acquisition. For simplicity and with some possible loss of performance, the raw measurement data can also be normalized according to the L2 norm of the central k-space signals. In our demonstration, we normalize the input data by the L2 norm of the central 5$\times$5 region of k-space. For intermediate layers, batch normalization can be used to minimize sensitivity to data scaling.

As training examples are difficult to obtain, the number of available training examples can be limited. To address this concern, data augmentation can be applied to train the reconstruction network. Care must be taken when applying data augmentation transformations. For example, data interpolation are needed for random rotations which may degrade the quality of the input data. Other image domain operations may introduce unrealistic errors in the measurement k-space domain, such as aliasing in the k-space domain. Flipping and transposing the dataset will preserve the original data quality; thus, these operations can be included in the training.

\subsection{Loss Function}

The performance of the data-driven approach is highly dependent on the loss function used. The easiest loss functions to use for training are L1 and L2 losses. The L2 loss is described in Eq.~\ref{eq:training} and can be converted to an L1 loss by using the L1 norm instead of the L2 norm. 

The L1 and L2 losses do not adequately capture the idea of structure or perception. More sophisticated loss functions can be used such as using a network pre-trained on natural images to extract ``perceptual'' features \cite{Johnson2016a}. Though general, this feature extraction network should be trained for the specific problem domain and the specific task at hand. Generative adversarial networks (GANs) \cite{Mardani2018} can be used to model the properties of the ground truth images and to exploit that information for improving the reconstruction quality. 

In Fig.~\ref{fig:network}C, we constructed a feature extraction network $D_\omega(m)$ that is trained to extract the necessary features to compare the reconstruction output with the ground truth \cite{Hammernik2018}. The training loss function in Eq.~\ref{eq:training} becomes:
\begin{equation}
    \min_\theta\max_\omega \sum_i \|D_\omega(G_\theta(y_i, A_i)) - D_\omega(m_i)\|_2,\label{eq:adversarial}
\end{equation}
where parameters $\omega$ are optimized to maximize the difference between the reconstructed image $G_\theta(y_i, A_i)$ and the ground truth data $m_i$. At the same time, parameters $\theta$ are optimized to minimize the difference between the reconstruction and the ground truth data after passing both these images through network $D_\omega$ to extract feature maps. The optimization of Eq.~\ref{eq:adversarial} consists of alternating between the training of parameters $\omega$ with $\theta$ constant and the training of parameters $\theta$ with $\omega$ constant. We refer to this training approach as training with an ``adversarial loss.''

This min-max loss function in Eq.~\ref{eq:adversarial} can be unstable and difficult to train. The many tricks used to train GANs, such as the training described in Ref.~\cite{Gulrajani2017}, can be leveraged here. We implement two main components to help stabilize the training \cite{Mardani2018}. First, $G_\theta$ is pre-trained using either an L1 or L2 loss so that the parameters in this network are properly initialized. The network is then fine tuned with the adversarial network using a reduced training rate. Second, a pixel-wise cost function is added to Eq.~\ref{eq:adversarial}:
\begin{equation}
    \min_\theta\max_\omega \sum_i \lambda\|D_\omega(G_\theta(y_i, A_i)) - D_\omega(m_i)\|_2^2 + \|G_\theta(y_i, A_i) - m_i\|_2^2.\label{eq:adversarial-l1}
\end{equation}
The images before passing through $D_\omega$ can be considered as additional feature maps to help stabilize the training process. Hyperparameter $\lambda$ is used to weigh between the two components in the loss function.

\begin{figure}
    \centering
    \includegraphics{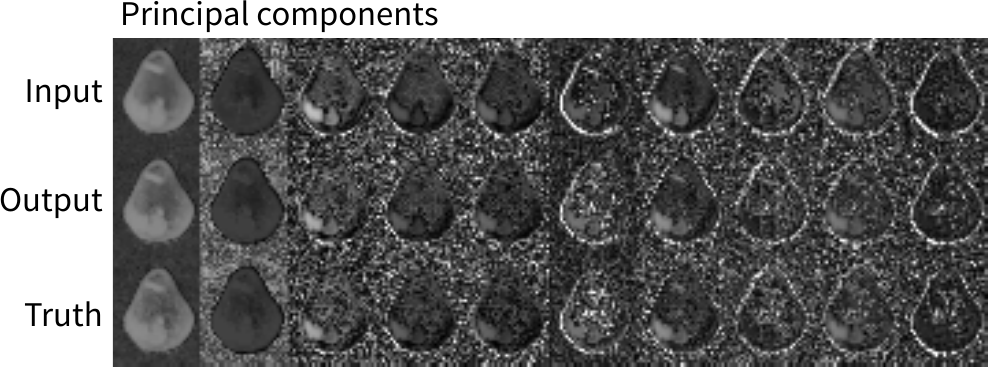}
    \caption{Example feature maps extracted by the learned network for adversarial loss. A feature extraction network was jointly trained with the reconstruction network. This feature extraction network was used to produce feature maps to compute the adversarial loss. Extracted features for the subsampled input (first row), reconstruction output (second row), and fully-sampled ground truth (last row) are displayed. The network extracted 128 different feature maps. Here, the dominant 10 principal components of the resulting feature maps are shown where all principal components are rotated to be aligned to the principal components of the truth data.}
    \label{fig:adversarial}
\end{figure}

\section{Demonstration}
The data should be collected at the point in the imaging pipeline where the model will be deployed. For the purpose of image reconstruction of subsampled datasets, we would need to collect the raw measurement k-space data. This data can be already filtered with the anti-aliasing readout filter and pre-whitened. Typically, this raw imaging data are not readily available as only magnitude images are saved as DICOM images in hospital imaging database. Furthermore, these stored magnitude images are often processed with image filters, and accurate simulation of raw imaging data from DICOM images is difficult to perform if not impossible especially in simulating realistic phase information. To facilitate development, a number of different open data initiatives have been recently launched, including \url{mridata.org}\cite{Ong2018a} and fastMRI \cite{Zbontar2018}. These resources provide an initial starting point for development, but more datasets of varying contrasts, field strengths, and vendors are needed.

For demonstration of the data-driven reconstruction and for enabling reproducibilty of the results, we downloaded 20 fully sampled volumetric knee datasets \cite{EppersonSMRT2013} that are freely available in the database of MRI raw data, \url{mridata.org} \cite{Ong2018a}. Each volume was collected with 320 slices in the readout direction in $x$, and each of these $x$-slices were treated as separate examples during training and validation. The datasets were divided by subject: 15 subjects for training (4800 $x$-slices), 2 subjects for validation (640 $x$-slices), and 3 subjects for testing. Sensitivity maps were estimated using JSENSE \cite{Ying2007}. Poisson-disc sampling masks \cite{Lustig2007a} were generated using an acceleration factor $R$ of 9.4 with corner cutting (effective $R$ of 12) and a fully-sampled calibration region of 20 $\times$ 20. 

\begin{figure}
    \centering
    \includegraphics[width=0.7\textwidth]{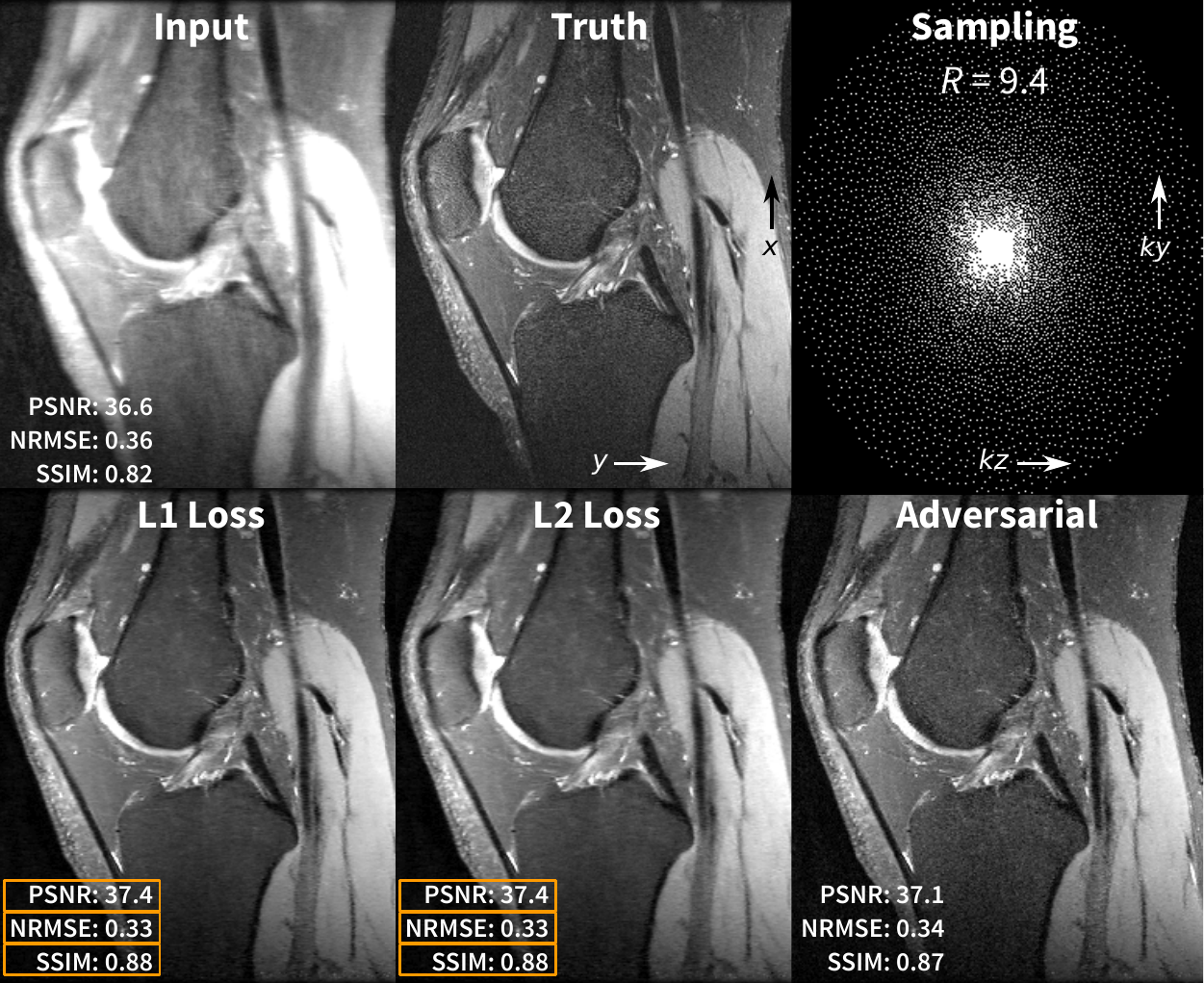}
    \caption{Demonstration results of a volumetric knee dataset that is subsampled with an acceleration factor $R$ of 9.4 with corner cutting (effective $R$ or 12). In the bottom row, the volume was reconstructed slice by slice using three networks trained with different loss functions. The reconstruction using the network trained using the L1 and L2 losses yielded the best results in terms of PSNR, NRMSE, and SSIM. However, the reconstruction using the network trained with the adversarial loss yielded results with most realistic texture.}
    \label{fig:demo:results}
\end{figure}

The networks in Fig.~\ref{fig:network} were implemented in python using the TensorFlow framework \cite{Abadi2016}. Additional reconstruction components were performed using the SigPy python package\footnote{\url{https://github.com/mikgroup/sigpy}} \cite{Ong2019}. The reconstruction network was built using 4 iterations; this setup allowed for relatively faster training for demonstration purposes. Performance can be improved by increasing the number of iterations. Training and experiments were performed on a single NVIDIA Titan Xp graphics card with 12GB of memory which supported a batch size of 2. The network was trained multiple times using different loss functions: an L1 loss for 20k steps and L2 loss for 20k steps. For adversarial loss, the network was first pre-trained using an L1 loss for 10k steps and then jointly trained with the adversarial loss and an L1 loss for 60k steps (10k steps were for the reconstruction network and 50k steps were for the adversarial feature extraction network). The data preparation, training, validation, and testing python scripts are available on GitHub. Example results are shown in Fig.~\ref{fig:demo:results}.

\section{Clinical Integration}

\begin{figure}
    \centering
    \includegraphics[width=0.4\textwidth]{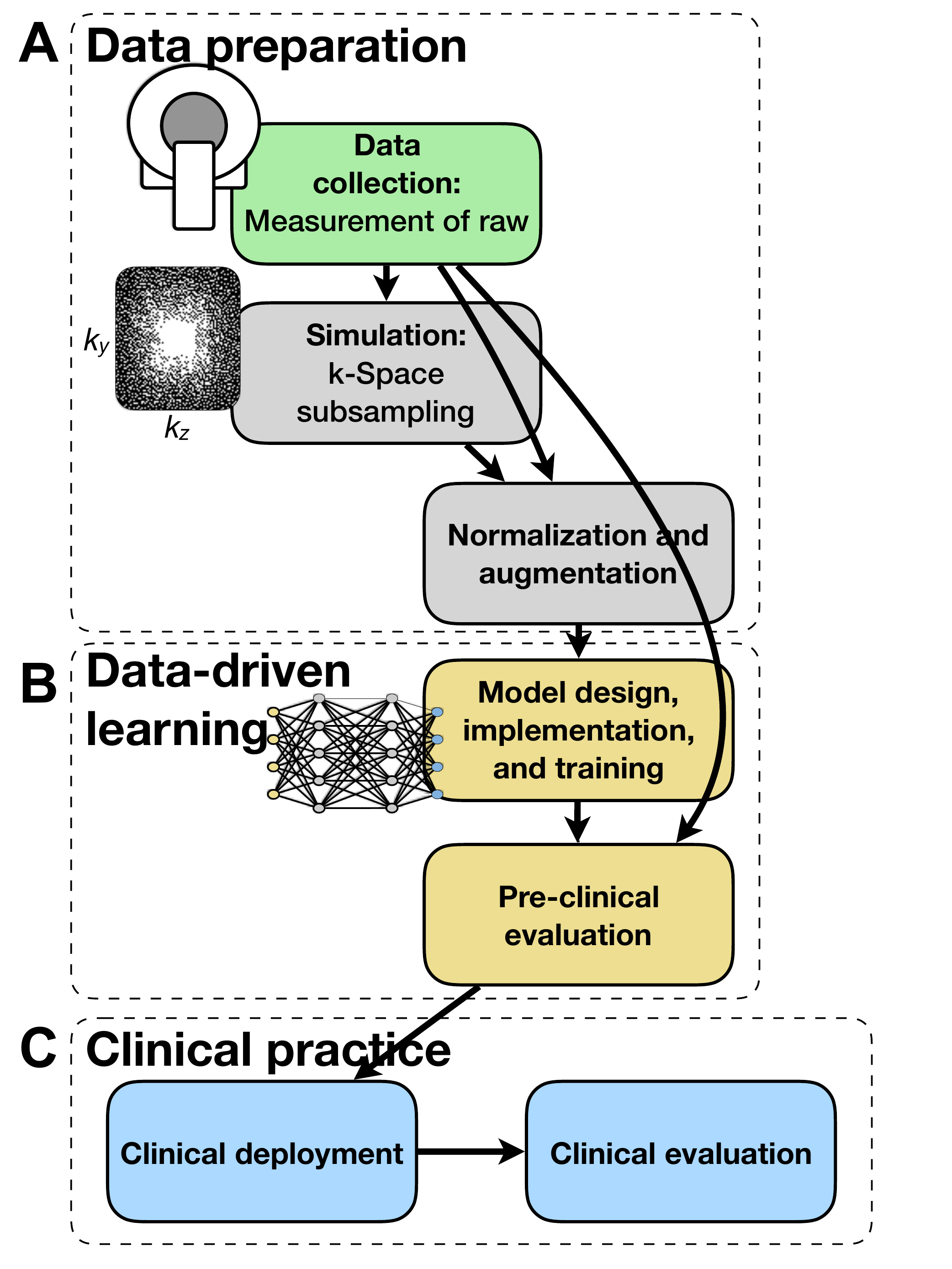}
    \caption{Example development cycle for the data-driven learning framework. In \textbf{A}, fully sampled raw k-space datasets are first collected and used to perform realistic simulation of subsampling. The reconstruction algorithm is then built and trained in \textbf{B}. Lastly, the reconstruction model is deployed and evaluated in \textbf{C}.}
    \label{fig:development}
\end{figure}

\subsection{Example Setup}
Many of these algorithms are computational intensive, but these algorithms can leverage off-the-shelf consumer hardware. The bulk of the computational burden is now in the network, and graphic processing units (GPUs) can be leveraged for this purpose. The compute system only needs the minimal central processing unit (CPU) and memory requirements to support the GPUs. Using an NVIDIA Titan Xp graphics card, one 320$\times$256 slice took on average 0.1 seconds to reconstruct, 1.8 seconds for a batch of 16 slices, and a total of 36.0 seconds for the entire volume with 320 slices. These benchmarks include the time to transfer the data to and from the CPU to the GPU. The computational speed can be further improved using newer GPU hardware and/or more cards. Alternatively, inference on the reconstruction network can be performed on the CPU to be able to leverage more memory in a more cost effective manner, but this setup comes with slower inference speed.

\subsection{Clinical Cases}
Through initial developments, we have deployed a number of different models at our clinical site. Example results of data-driven compressed sensing with variational networks (VN) \cite{Hammernik2017,Chen2018b} are shown in Fig.~\ref{fig:demo:cs_reg}. Conventional compressed sensing using L1-ESPIRiT \cite{Uecker2013} required tuning of the regularization parameter $\lambda$. Conventional compressed sensing reconstruction achieved high noise when the regularization parameter was too low (0.0005, left column in Fig.~\ref{fig:demo:cs_reg}), and high residual and blurring artifacts when the regularization parameter was too high (0.05, right column in Fig.~\ref{fig:demo:cs_reg}). Optimal value of this parameter may vary with different scans. Compared to conventional compressed sensing, learning a VN (middle column in Fig.~\ref{fig:demo:cs_reg}) achieved proper regularization without the need of tuning the regularization parameter.

\begin{figure}
    \centering
    \includegraphics{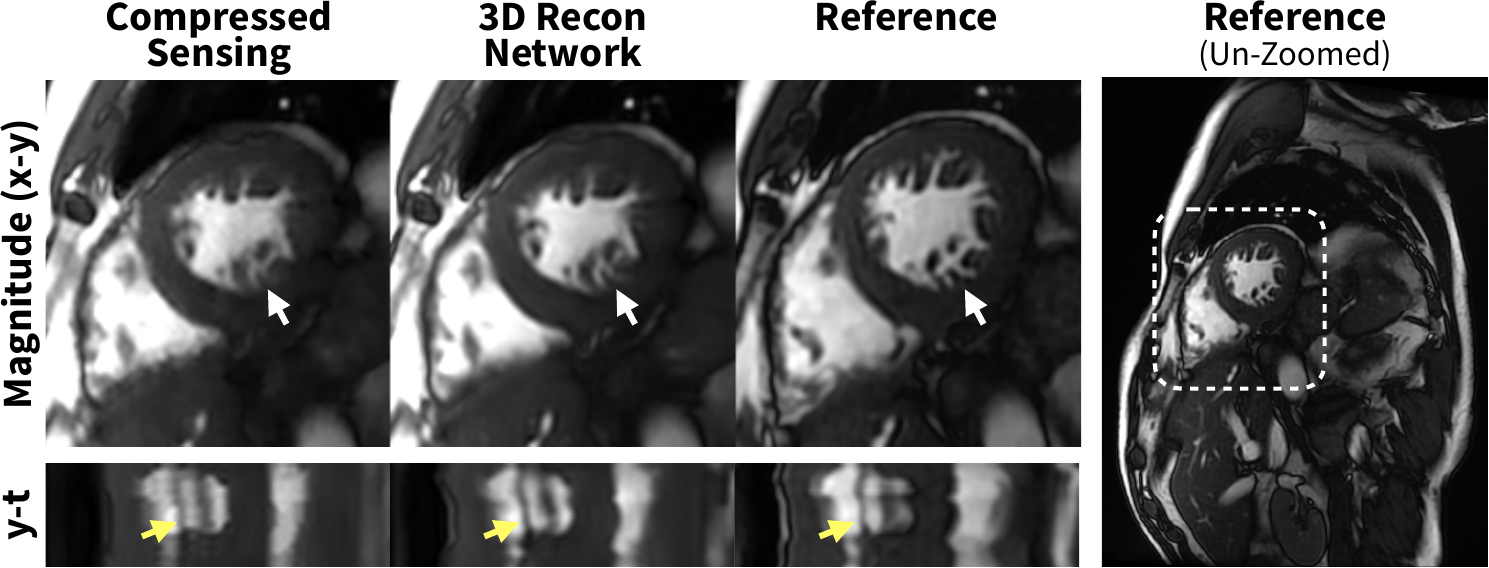}
    \caption{Clinical deployment of a 3D spatiotemporal convolutional reconstruction network for two-dimensional cardiac-resolved imaging. Above, a prospectively acquired short-axis dataset with 12.6 fold subsampling was reconstructed using compressed sensing (L1-ESPIRiT) \cite{Uecker2013} and a trained 3D reconstruction network. These reconstructions were compared with the fully sampled reference images (acquired in a separate scan). The 3D network was able to reconstruct fine structures such as papillary muscles and trabeculae with less blurring than L1-ESPIRiT (white arrows). Furthermore, the images from the network depict more natural cardiac motion compared to L1-ESPIRiT with total variation, which introduced block artifacts (yellow arrows).}
    \label{fig:cine}
\end{figure}

The best image prior for multi-dimensional imaging is more difficult to engineer, but these larger datasets benefits tremendously from subsampling since these datasets take longer to acquire. Thus, for multi-dimensional space, a compelling approach is to use data-driven learning. In Fig.~\ref{fig:cine}, a reconstruction network consisting of 3D spatiotemporal convolutions was trained on 12 fully-sampled, breath-held, multi-slice cardiac cine datasets acquired with a balanced steady-state free precession (bSSFP) readout at multiple scan orientations. The reconstruction network enabled the acquisition of a cine slice in a single heartbeat and a full stack of cine slices in a single breathhold \cite{Sandino2019}.

\subsection{Evaluation}
For development and prototyping, image metrics provide a straightforward method to evaluate performance. These metrics include peak signal-to-noise ratio (PSNR), normalized root-mean-square error (NRMSE), and structural similarity (SSIM) \cite{Wang2004d}. These metrics are described in the Appendix. However, what is more important in the clinical setting is the ability for clinicians to make an informed decision. Therefore, we recommend performing clinical studies. 
For comparing methods, the evaluation of clinical studies can be performed with blinded grading of image quality by multiple radiologists. This evaluation can be based on set criteria including overall image quality, perceived signal-to-noise ratio, image contrast, image sharpness, residual artifacts, and confidence of diagnosis \cite{Chen2018b}. Scores from 1 to 5 or -2 to 2 are given for each clinical patient scan with a specific definition for each score for objectivity. An example of scoring for signal-to-noise ratio and image sharpness is described in Table \ref{tab:eval}. 

A total number of more than 20 consecutive scans are usually collected to achieve a comprehensive and statistically meaningful evaluation for an initial study. The difference between two reconstruction approaches can be statistically tested with Wilcoxon tests on the null hypothesis that there was no significant difference between two approaches.

\begin{table}[h]
    \centering
    \small
    \begin{tabular}{clll}
        \hline\hline
        Score &  Overall Image Quality & Signal-to-Noise Ratio & Sharpness\\
        \hline
        1 & Nondiagnostic & 
        \begin{tabular}{@{}l@{}}
            All structures appear to be\\ 
            \hspace{5mm}too noisy.
        \end{tabular} &
        \begin{tabular}{@{}l@{}}
            Some structures are not\\
            \hspace{5mm}sharp on most images.
        \end{tabular}\\
        2 & Limited & 
        \begin{tabular}{@{}l@{}}
            Most structures appear to be\\
            \hspace{5mm}too noisy.
        \end{tabular} &
        \begin{tabular}{@{}l@{}}
            Most structures are sharp\\
            \hspace{5mm}on some images.
        \end{tabular}\\
        3 & Diagnostic & 
        \begin{tabular}{@{}l@{}}
            Few structures appear to be\\ 
            \hspace{5mm}too noisy on most images.
        \end{tabular} &
        \begin{tabular}{@{}l@{}}
            Most structures are sharp\\
            \hspace{5mm}on most images.
        \end{tabular}\\
        4 & Good & 
        \begin{tabular}{@{}l@{}}
            Few structures appear to be\\
            \hspace{5mm}too noisy on a few images.
        \end{tabular} &
        \begin{tabular}{@{}l@{}}
            All structures are sharp\\
            \hspace{5mm}on most images.\\
        \end{tabular}\\
        5 & Excellent & 
        \begin{tabular}{@{}l@{}}
            There is no noticeable noise\\
            \hspace{5mm}on any of the images.
        \end{tabular} &
        \begin{tabular}{@{}l@{}}
            All structures are sharp\\
            \hspace{5mm}on all images.\\
        \end{tabular}\\
        \hline\hline
    \end{tabular}
    \caption{Example scoring criteria to evaluate methods.}
    \label{tab:eval}
\end{table}

\section{Discussion}
This data-driven approach to accelerated imaging has the potential to eliminate many of the challenges associated with compressed sensing, including the need to design hand-crafted priors and to hand tune the regularization functions. In addition, the described framework inspired by iterative inference algorithms provides a principled approach for designing reconstruction networks. Furthermore, this approach is useful to interpret the trained network components. However, the data-driven reconstruction framework faces new challenges, including data scarcity, generalizability, reliability, and meaningful metrics. 

High quality training labels, or fully sampled datasets, are scarce especially in the clinical settings where patient motion impacts image quality and where lengthy fully-sampled acquisitions are impractical to perform if a faster solution exists. To address data scarcity, an important future direction pertains to designing compact network architectures that are effective with small training labels and are possibly trained in an unsupervised fashion. Early attempts are made in \cite{Mardani2018a} where recurrent neural networks are leveraged for learning proximal operators using only a couple of residual blocks that performs well for small training sample sizes. 

The network can be trained or re-trained for different anatomies and different types of scan contrasts. This strategy can be implemented if sufficient training datasets for all different settings are readily available. Given the scarcity of training examples and the cost to collect these examples, another strategy is to design and train networks that are highly generalizable. This generalizability can be achieved with smaller networks as discussed before or by training with a larger image manifold such as using natural images. With this approach, a loss in performance may be observed since a larger than necessary manifold is learned. In the worst case, the images reconstructed using this data-driven approach should not be worse than images reconstructed using a more conservative approach such as zero-filling or parallel imaging \cite{Pruessmann1999,Uecker2013}.

Regarding generalizability, an important question concerns reconstruction performance for subjects with unseen abnormalities. Patient abnormalities can be quite heterogeneous, and these abnormalities are rare and unlikely to be included in the training dataset. If not designed carefully, generative networks have the possibility of removing or creating critical features that will result in misdiagnosis. The optimization-based network architecture utilizes data consistency, as exemplified by the gradient update step, for the image recovery problem. However, the inherent ambiguity of ill-posed problems does not guarantee faithful recovery. Therefore, a systematic study is required to analyze the recovery of images. Also, effective regularization techniques (possibly through adversarial training) are needed to avoid missing important diagnostic information. More efforts are needed to develop a holistic quality score capturing the uncertainty in the acquisition scheme and training data. 


Developing a standard unbiased metric for medical images that assesses the authenticity of medical images is extremely important. Here, we discuss the use of different loss functions to train the network and common imaging metrics to evaluate the images. Additionally, we discuss an example of a possible clinical evaluation that can be performed to assess the algorithm in the clinical setting. However, the reconstruction task should ideally be optimized and evaluated for the end task which can be consisted of detection and quantification. With significant effort in automating image interpretation, this data-driven framework provides an opportunity to pursue the ability to train the reconstruction end-to-end for the ultimate goal of improving patient care. 

In conclusion, deep learning has the potential to increase the accessibility and generalizability of fast imaging through data subsampling. Previous challenges with compressed sensing can be approached with a data-driven framework to create a solution that is more readily translated to clinical practice. 

\appendix
\section*{Appendix: Imaging Metrics}
Imaging metrics are commonly used to evaluate results. This includes peak signal-to-noise ratio (PSNR) and normalized root-mean-square error (NRMSE). For these quantities, we use the following equations:
\begin{align}
    \mathrm{MSE}(x, x_{r}) &= \frac{1}{NM}\sum_i^N\sum_j^M |x[i,j] - x_{r}[i,j]|^2,\\
    \mathrm{PSNR}(x, x_{r}) &= 10 \log_{10}\left( \max(|x_r|^2) / \mathrm{MSE}\right),\\
    \mathrm{NRMSE}(x, x_{r}) &= \sqrt{\mathrm{MSE}} / \sqrt{\frac{1}{NM}\sum_i^N\sum_j^M |x_{r}[i,j]|^2},
\end{align}
where $x$ denotes the test image, $x[i,j]$ denotes the value of the test image at pixel $(i, j)$, and $x_r$ denotes the reference ground truth image.

\section*{Acknowledgment}
The authors would like to thank GE Healthcare, NIH R01-EB009690, NIH R01-EB026136, and NIH R01-EB019241 for the research support.

\bibliographystyle{IEEEtran}
\bibliography{jycheng,feiyu}

\begin{thebibliography}{10}
\providecommand{\url}[1]{#1}
\csname url@samestyle\endcsname
\providecommand{\newblock}{\relax}
\providecommand{\bibinfo}[2]{#2}
\providecommand{\BIBentrySTDinterwordspacing}{\spaceskip=0pt\relax}
\providecommand{\BIBentryALTinterwordstretchfactor}{4}
\providecommand{\BIBentryALTinterwordspacing}{\spaceskip=\fontdimen2\font plus
\BIBentryALTinterwordstretchfactor\fontdimen3\font minus
  \fontdimen4\font\relax}
\providecommand{\BIBforeignlanguage}[2]{{%
\expandafter\ifx\csname l@#1\endcsname\relax
\typeout{** WARNING: IEEEtran.bst: No hyphenation pattern has been}%
\typeout{** loaded for the language `#1'. Using the pattern for}%
\typeout{** the default language instead.}%
\else
\language=\csname l@#1\endcsname
\fi
#2}}
\providecommand{\BIBdecl}{\relax}
\BIBdecl

\bibitem{Pruessmann1999}
\BIBentryALTinterwordspacing
K.~P. Pruessmann, M.~Weiger, M.~B. Scheidegger, and P.~Boesiger, ``{SENSE:
  sensitivity encoding for fast MRI.}'' \emph{Magnetic Resonance in Medicine},
  vol.~42, no.~5, pp. 952--62, 11 1999. [Online]. Available:
  \url{http://www.ncbi.nlm.nih.gov/pubmed/10542355}
\BIBentrySTDinterwordspacing

\bibitem{Ying2007}
\BIBentryALTinterwordspacing
L.~Ying and J.~Sheng, ``{Joint image reconstruction and sensitivity estimation
  in SENSE (JSENSE)},'' \emph{Magnetic Resonance in Medicine}, vol.~57, no.~6,
  pp. 1196--1202, 6 2007. [Online]. Available:
  \url{http://doi.wiley.com/10.1002/mrm.21245}
\BIBentrySTDinterwordspacing

\bibitem{Uecker2013}
\BIBentryALTinterwordspacing
M.~Uecker, P.~Lai, M.~J. Murphy, P.~Virtue, M.~Elad, J.~M. Pauly, S.~S.
  Vasanawala, and M.~Lustig, ``{ESPIRiT-an eigenvalue approach to
  autocalibrating parallel MRI: Where SENSE meets GRAPPA},'' \emph{Magnetic
  Resonance in Medicine}, vol.~71, no.~3, pp. 990--1001, 3 2014. [Online].
  Available: \url{http://doi.wiley.com/10.1002/mrm.24751}
\BIBentrySTDinterwordspacing

\bibitem{Lustig2007a}
\BIBentryALTinterwordspacing
M.~Lustig, D.~Donoho, and J.~M. Pauly, ``{Sparse MRI: The application of
  compressed sensing for rapid MR imaging.}'' \emph{Magnetic Resonance in
  Medicine}, vol.~58, no.~6, pp. 1182--1195, 12 2007. [Online]. Available:
  \url{http://www.ncbi.nlm.nih.gov/pubmed/17969013}
\BIBentrySTDinterwordspacing

\bibitem{Cheng2015a}
\BIBentryALTinterwordspacing
J.~Y. Cheng, K.~Hanneman, T.~Zhang, M.~T. Alley, P.~Lai, J.~I. Tamir,
  M.~Uecker, J.~M. Pauly, M.~Lustig, and S.~S. Vasanawala, ``{Comprehensive
  motion-compensated highly accelerated 4D flow MRI with ferumoxytol
  enhancement for pediatric congenital heart disease.}'' \emph{Journal of
  Magnetic Resonance Imaging}, vol.~43, no.~6, pp. 1355--1368, 6 2016.
  [Online]. Available: \url{http://www.ncbi.nlm.nih.gov/pubmed/26646061}
\BIBentrySTDinterwordspacing

\bibitem{Vasanawala2015}
\BIBentryALTinterwordspacing
S.~S. Vasanawala, K.~Hanneman, M.~T. Alley, and A.~Hsiao, ``{Congenital heart
  disease assessment with 4D flow MRI.}'' \emph{Magnetic Resonance in
  Medicine}, vol.~42, no.~4, pp. 870--86, 10 2015. [Online]. Available:
  \url{http://www.ncbi.nlm.nih.gov/pubmed/25708923}
\BIBentrySTDinterwordspacing

\bibitem{jung2009k}
H.~Jung, K.~Sung, K.~S. Nayak, E.~Y. Kim, and J.~C. Ye, ``k-t focuss: a general
  compressed sensing framework for high resolution dynamic mri,''
  \emph{Magnetic resonance in medicine}, vol.~61, no.~1, pp. 103--116, 2009.

\bibitem{Feng2012}
\BIBentryALTinterwordspacing
L.~Feng, M.~B. Srichai, R.~P. Lim, A.~Harrison, W.~King, G.~Adluru, E.~V.~R.
  Dibella, D.~K. Sodickson, R.~Otazo, and D.~Kim, ``{Highly accelerated
  real-time cardiac cine MRI using k-t SPARSE-SENSE.}'' \emph{Magnetic
  Resonance in Medicine}, vol.~70, no.~1, pp. 64--74, 7 2013. [Online].
  Available: \url{http://www.ncbi.nlm.nih.gov/pubmed/22887290}
\BIBentrySTDinterwordspacing

\bibitem{Lingala2011}
\BIBentryALTinterwordspacing
S.~G. Lingala, Y.~Hu, E.~DiBella, and M.~Jacob, ``{Accelerated dynamic MRI
  exploiting sparsity and low-rank structure: k-t SLR.}'' \emph{IEEE
  Transactions on Medical Imaging}, vol.~30, no.~5, pp. 1042--1054, 5 2011.
  [Online]. Available: \url{http://www.ncbi.nlm.nih.gov/pubmed/21292593}
\BIBentrySTDinterwordspacing

\bibitem{Zhang2013a}
\BIBentryALTinterwordspacing
T.~Zhang, J.~Y. Cheng, A.~G. Potnick, R.~A. Barth, M.~T. Alley, M.~Uecker,
  M.~Lustig, J.~M. Pauly, and S.~S. Vasanawala, ``{Fast pediatric 3D
  free-breathing abdominal dynamic contrast enhanced MRI with high
  spatiotemporal resolution},'' \emph{Journal of Magnetic Resonance Imaging},
  vol.~41, no.~2, pp. 460--473, 2 2015. [Online]. Available:
  \url{http://www.ncbi.nlm.nih.gov/pubmed/24375859}
\BIBentrySTDinterwordspacing

\bibitem{Feng2015}
\BIBentryALTinterwordspacing
L.~Feng, L.~Axel, H.~Chandarana, K.~T. Block, D.~K. Sodickson, and R.~Otazo,
  ``{XD-GRASP: Golden-angle radial MRI with reconstruction of extra
  motion-state dimensions using compressed sensing.}'' \emph{Magnetic Resonance
  in Medicine}, vol.~75, no.~2, pp. 775--88, 2 2016. [Online]. Available:
  \url{http://www.ncbi.nlm.nih.gov/pubmed/25809847}
\BIBentrySTDinterwordspacing

\bibitem{Cheng2017ScientificReports}
\BIBentryALTinterwordspacing
J.~Y. Cheng, T.~Zhang, M.~T. Alley, M.~Uecker, M.~Lustig, J.~M. Pauly, and
  S.~S. Vasanawala, ``{Comprehensive Multi-Dimensional MRI for the Simultaneous
  Assessment of Cardiopulmonary Anatomy and Physiology},'' \emph{Scientific
  Reports}, vol.~7, no.~1, p. 5330, 2017. [Online]. Available:
  \url{http://dx.doi.org/10.1038/s41598-017-04676-8}
\BIBentrySTDinterwordspacing

\bibitem{Vasanawala2010a}
\BIBentryALTinterwordspacing
S.~S. Vasanawala, M.~T. Alley, B.~A. Hargreaves, R.~A. Barth, J.~M. Pauly, and
  M.~Lustig, ``{Improved Pediatric MR Imaging with Compressed Sensing},''
  \emph{Radiology}, vol. 256, no.~2, pp. 607--616, 8 2010. [Online]. Available:
  \url{http://www.ncbi.nlm.nih.gov/pubmed/20529991}
\BIBentrySTDinterwordspacing

\bibitem{Hammernik2017}
K.~Hammernik, T.~Klatzer, E.~Kobler, M.~P. Recht, D.~K. Sodickson, T.~Pock, and
  F.~Knoll, ``{Learning a variational network for reconstruction of accelerated
  MRI data},'' \emph{Magnetic Resonance in Medicine}, vol.~79, no.~6, pp.
  3055--3071, 6 2018.

\bibitem{Chen2018b}
\BIBentryALTinterwordspacing
F.~Chen, V.~Taviani, I.~Malkiel, J.~Y. Cheng, J.~I. Tamir, J.~Shaikh, S.~T.
  Chang, C.~J. Hardy, J.~M. Pauly, and S.~S. Vasanawala, ``{Variable-Density
  Single-Shot Fast Spin-Echo MRI with Deep Learning Reconstruction by Using
  Variational Networks},'' \emph{Radiology}, vol. 289, no.~2, pp. 366--373, 11
  2018. [Online]. Available:
  \url{http://pubs.rsna.org/doi/10.1148/radiol.2018180445}
\BIBentrySTDinterwordspacing

\bibitem{ramani2012regularization}
S.~Ramani, Z.~Liu, J.~Rosen, J.-F. Nielsen, and J.~A. Fessler, ``Regularization
  parameter selection for nonlinear iterative image restoration and mri
  reconstruction using gcv and sure-based methods,'' \emph{IEEE Transactions on
  Image Processing}, vol.~21, no.~8, pp. 3659--3672, 2012.

\bibitem{Beck2009}
\BIBentryALTinterwordspacing
A.~Beck and M.~Teboulle, ``{A fast iterative shrinkage-thresholding algorithm
  for linear inverse problems},'' \emph{SIAM J Imaging Sci}, vol.~2, no.~1, pp.
  183--202, 2009. [Online]. Available:
  \url{http://epubs.siam.org/doi/pdf/10.1137/080716542}
\BIBentrySTDinterwordspacing

\bibitem{Yang2016a}
\BIBentryALTinterwordspacing
Y.~Yang, J.~Sun, H.~Li, and Z.~Xu, ``{ADMM-Net: A Deep Learning Approach for
  Compressive Sensing MRI},'' \emph{NIPS}, pp. 10--18, 5 2017. [Online].
  Available: \url{http://arxiv.org/abs/1705.06869}
\BIBentrySTDinterwordspacing

\bibitem{Adler2017}
\BIBentryALTinterwordspacing
J.~Adler and O.~Oktem, ``{Learned Primal-Dual Reconstruction},'' \emph{IEEE
  Transactions on Medical Imaging}, vol.~37, no.~6, pp. 1322--1332, 6 2018.
  [Online]. Available: \url{https://ieeexplore.ieee.org/document/8271999/}
\BIBentrySTDinterwordspacing

\bibitem{aggarwal2018modl}
H.~K. Aggarwal, M.~P. Mani, and M.~Jacob, ``Modl: Model based deep learning
  architecture for inverse problems,'' \emph{IEEE transactions on medical
  imaging}, 2018.

\bibitem{Cheng2018}
\BIBentryALTinterwordspacing
J.~Y. Cheng, F.~Chen, M.~T. Alley, J.~M. Pauly, and S.~S. Vasanawala, ``{Highly
  Scalable Image Reconstruction using Deep Neural Networks with Bandpass
  Filtering},'' \emph{arXiv:1805.03300 [cs.CV]}, 5 2018. [Online]. Available:
  \url{http://arxiv.org/abs/1805.03300}
\BIBentrySTDinterwordspacing

\bibitem{Daubechies2004a}
\BIBentryALTinterwordspacing
I.~Daubechies, M.~Defrise, and C.~De~Mol, ``{An iterative thresholding
  algorithm for linear inverse problems with a sparsity constraint},''
  \emph{Communications on Pure and Applied Mathematics}, vol.~57, no.~11, pp.
  1413--1457, 11 2004. [Online]. Available:
  \url{http://doi.wiley.com/10.1002/cpa.20042}
\BIBentrySTDinterwordspacing

\bibitem{He2016a}
\BIBentryALTinterwordspacing
K.~He, X.~Zhang, S.~Ren, and J.~Sun, ``{Identity Mappings in Deep Residual
  Networks},'' \emph{arXiv:1603.05027 [cs.CV]}, 3 2016. [Online]. Available:
  \url{http://arxiv.org/abs/1603.05027}
\BIBentrySTDinterwordspacing

\bibitem{Trabelsi2017}
\BIBentryALTinterwordspacing
C.~Trabelsi, O.~Bilaniuk, Y.~Zhang, D.~Serdyuk, S.~Subramanian, J.~F. Santos,
  S.~Mehri, N.~Rostamzadeh, Y.~Bengio, and C.~J. Pal, ``{Deep Complex
  Networks},'' \emph{arXiv:1705.09792 [cs.NE]}, pp. 1--19, 5 2017. [Online].
  Available: \url{http://arxiv.org/abs/1705.09792}
\BIBentrySTDinterwordspacing

\bibitem{Johnson2016a}
\BIBentryALTinterwordspacing
J.~Johnson, A.~Alahi, and L.~Fei-Fei, ``{Perceptual Losses for Real-Time Style
  Transfer and Super-Resolution},'' in \emph{ECCV}, 3 2016. [Online].
  Available: \url{http://arxiv.org/abs/1603.08155}
\BIBentrySTDinterwordspacing

\bibitem{Mardani2018}
\BIBentryALTinterwordspacing
M.~Mardani, E.~Gong, J.~Y. Cheng, S.~S. Vasanawala, G.~Zaharchuk, L.~Xing, and
  J.~M. Pauly, ``{Deep Generative Adversarial Neural Networks for Compressive
  Sensing MRI},'' \emph{IEEE Transactions on Medical Imaging}, vol.~38, no.~1,
  pp. 167--179, 1 2019. [Online]. Available:
  \url{https://ieeexplore.ieee.org/document/8417964/}
\BIBentrySTDinterwordspacing

\bibitem{Hammernik2018}
K.~Hammernik, E.~Kobler, T.~Pock, M.~P. Recht, D.~K. Sodickson, and F.~Knoll,
  ``{Variational Adversarial Networks for Accelerated MR Image
  Reconstruction},'' in \emph{Joint Annual Meeting ISMRM-ESMRMB}, Paris,
  France, 2018, p. 1091.

\bibitem{Gulrajani2017}
\BIBentryALTinterwordspacing
I.~Gulrajani, F.~Ahmed, M.~Arjovsky, V.~Dumoulin, and A.~Courville, ``{Improved
  Training of Wasserstein GANs},'' \emph{arXiv:1704.00028}, 2017. [Online].
  Available: \url{http://arxiv.org/abs/1704.00028}
\BIBentrySTDinterwordspacing

\bibitem{Ong2018a}
F.~Ong, S.~Amin, S.~S. Vasanawala, and M.~Lustig, ``{An Open Archive for
  Sharing MRI Raw Data},'' in \emph{ISMRM {\&} ESMRMB Joint Annual Meeting},
  Paris, France, 2018, p. 3425.

\bibitem{Zbontar2018}
J.~Zbontar, F.~Knoll, A.~Sriram, M.~J. Muckley, M.~Bruno, A.~Defazio,
  M.~Parente, K.~J. Geras, J.~Katsnelson, H.~Chandarana, Z.~Zhang, M.~Drozdzal,
  A.~Romero, M.~Rabbat, P.~Vincent, J.~Pinkerton, D.~Wang, N.~Yakubova,
  E.~Owens, C.~L. Zitnick, M.~P. Recht, D.~K. Sodickson, and Y.~W. Lui,
  ``{fastMRI : An Open Dataset and Benchmarks for Accelerated MRI},''
  \emph{arXiv: 11811.08839 [cs.CV]}, pp. 1--29, 2018.

\bibitem{EppersonSMRT2013}
K.~Epperson, A.~M. Sawyer, M.~Lustig, M.~T. Alley, M.~Uecker, P.~Virtue,
  P.~Lai, and S.~S. Vasanawala, ``{Creation of Fully Sampled MR Data Repository
  for Compressed Sensing of the Knee},'' in \emph{SMRT 22nd Annual Meeting},
  Salt Lake City, Utah, USA, 2013.

\bibitem{Abadi2016}
\BIBentryALTinterwordspacing
M.~Abadi, A.~Agarwal, P.~Barham, E.~Brevdo, Z.~Chen, C.~Citro, G.~S. Corrado,
  A.~Davis, J.~Dean, M.~Devin, S.~Ghemawat, I.~Goodfellow, A.~Harp, G.~Irving,
  M.~Isard, Y.~Jia, R.~Jozefowicz, L.~Kaiser, M.~Kudlur, J.~Levenberg, D.~Mane,
  R.~Monga, S.~Moore, D.~Murray, C.~Olah, M.~Schuster, J.~Shlens, B.~Steiner,
  I.~Sutskever, K.~Talwar, P.~Tucker, V.~Vanhoucke, V.~Vasudevan, F.~Viegas,
  O.~Vinyals, P.~Warden, M.~Wattenberg, M.~Wicke, Y.~Yu, and X.~Zheng,
  ``{TensorFlow: Large-Scale Machine Learning on Heterogeneous Distributed
  Systems},'' \emph{arXiv:1603.04467 [cs.DC]}, 3 2016. [Online]. Available:
  \url{http://arxiv.org/abs/1603.04467}
\BIBentrySTDinterwordspacing

\bibitem{Ong2019}
F.~Ong and M.~Lustig, ``{SigPy: A Python Package for High Performance Iterative
  Reconstruction},'' in \emph{ISMRM Annual Meeting {\&} Exhibition}, Montreal,
  Canada, 2019.

\bibitem{Sandino2019}
C.~Sandino, P.~Lai, M.~A. Janich, A.~C.~S. Brau, S.~S. Vasanawala, and J.~Y.
  Cheng, ``{ESPIRiT with deep priors: Accelerating 2D cardiac CINE MRI beyond
  compressed sensing},'' in \emph{SCMR/ISMRM Workshop}, Seattle, Washington,
  USA, 2019.

\bibitem{Wang2004d}
\BIBentryALTinterwordspacing
Z.~Wang, A.~C. Bovik, H.~R. Sheikh, and E.~P. Simoncelli, ``{Image quality
  assessment: from error visibility to structural similarity.}'' \emph{IEEE
  Trans Image Processing}, vol.~13, no.~4, pp. 600--612, 4 2004. [Online].
  Available: \url{http://www.ncbi.nlm.nih.gov/pubmed/15376593}
\BIBentrySTDinterwordspacing

\bibitem{Mardani2018a}
\BIBentryALTinterwordspacing
M.~Mardani, Q.~Sun, S.~Vasawanala, V.~Papyan, H.~Monajemi, J.~Pauly, and
  D.~Donoho, ``{Neural Proximal Gradient Descent for Compressive Imaging},''
  \emph{Conference on Neural Information Processing Systems}, vol. 8166, p.
  1–11, 2018. [Online]. Available: \url{http://arxiv.org/abs/1806.03963}
\BIBentrySTDinterwordspacing

\end{thebibliography}

\end{document}